\documentclass[sigconf, nonacm, pdfa]{acmart}

\usepackage[a-2b]{pdfx}

\usepackage{graphicx}
\usepackage{draftfigure}

\usepackage{multirow}
\usepackage{booktabs}
\usepackage{colortbl}
\usepackage{makecell}
\usepackage{threeparttable}
\usepackage{tabu}
\usepackage{boldline}

\usepackage{algorithmic}
\usepackage{textcomp}

\usepackage{fontawesome}
\usepackage{pifont}
\usepackage{utfsym}
\usepackage{marvosym}

\definecolor{babyblueeyes}{rgb}{0.63, 0.79, 0.95}
\definecolor{gray(x11gray)}{rgb}{0.75, 0.75, 0.75}
\definecolor{gray(html/cssgray)}{rgb}{0.5, 0.5, 0.5}
\definecolor{green(munsell)}{rgb}{0.0, 0.66, 0.47}

\definecolor{dbblue}{RGB}{200,230,255}
\definecolor{lmgreen}{RGB}{230,255,230}

\newcommand{\strong}{\textcolor{green}{$\checkmark$}}
\newcommand{\weak}{\textcolor{red}{$\times$}}
\newcommand{\moderate}{\textcolor{blue}{$\approx$}}

\newcommand{\method}{\texttt}

\newcommand\vldbyear{2025}
\newcommand\vldbworkshop{Governance, Understanding and Integration of Data for Effective and Responsible AI (GUIDE-AI '25)}
\newcommand\vldbauthors{\authors}
\newcommand\vldbtitle{\shorttitle} 
\newcommand\vldbavailabilityurl{}
\newcommand\vldbpagestyle{empty} 

\begin{document}
\title{DBMS-LLM Integration Strategies in Industrial and Business Applications: Current Status and Future Challenges}

\author{Zhengtong Yan}
\affiliation{%
  \institution{University of Helsinki}
  \city{Helsinki}
  \country{Finland}
}
\email{zhengtong.yan@helsinki.fi}

\author{Gongsheng Yuan}
\affiliation{%
  \institution{Zhejiang University}
  \city{Hangzhou}
  \country{China}
}
\email{ygs@zju.edu.cn}

\author{Qingsong Guo}
\affiliation{
  \institution{Hunan University of Technology}
  \city{Zhuzhou}
  \country{China}
}
\email{qingsongg@gmail.com}

\author{Jiaheng Lu}
\affiliation{%
  \institution{University of Helsinki}
  \city{Helsinki}
  \country{Finland}
}
\email{jiaheng.lu@helsinki.fi}

\begin{abstract}

Modern enterprises are increasingly driven by the \emph{DATA+AI} paradigm, in which Database Management Systems (DBMSs) and Large Language Models (LLMs) have become two foundational infrastructures powering a wide range of industrial and business applications, such as enterprise analytics, intelligent customer service, and data-driven decision-making. The efficient integration of DBMSs and LLMs within a unified system offers significant opportunities but also introduces new technical challenges. This paper surveys recent developments in DBMS–LLM integration and identifies key future challenges. Specifically, we categorize five representative architectural patterns based on their core design principles, strengths, and trade-offs. Based on this analysis, we further highlight several critical open challenges. We aim to provide a systematic understanding of the current integration landscape and to outline the unresolved issues that must be addressed to achieve scalable and efficient integration of traditional data management and advanced language reasoning in future intelligent applications.

\end{abstract}

\maketitle

\pagestyle{\vldbpagestyle}
\begingroup\small\noindent\raggedright\textbf{VLDB Workshop Reference Format:}\\
\vldbauthors. \vldbtitle. VLDB \vldbyear\ Workshop: \vldbworkshop.\\ 
\endgroup
\begingroup
\renewcommand\thefootnote{}\footnote{\noindent
This work is licensed under the Creative Commons BY-NC-ND 4.0 International License. Visit \url{https://creativecommons.org/licenses/by-nc-nd/4.0/} to view a copy of this license. For any use beyond those covered by this license, obtain permission by emailing \href{mailto:info@vldb.org}{info@vldb.org}. Copyright is held by the owner/author(s). Publication rights licensed to the VLDB Endowment. \\
\raggedright Proceedings of the VLDB Endowment. 
ISSN 2150-8097. \\
}\addtocounter{footnote}{-1}\endgroup

\ifdefempty{\vldbavailabilityurl}{}{
\vspace{.3cm}
\begingroup\small\noindent\raggedright\textbf{VLDB Workshop Artifact Availability:}\\
The source code, data, and/or other artifacts have been made available at \url{\vldbavailabilityurl}.
\endgroup
}

\section{Introduction}\label{sec:intro}
Database Management Systems (DBMSs) have served as the foundational infrastructure of modern enterprises since the 1970s~\cite{codd1970relational}. Over the past decades, DBMSs have evolved significantly across multiple dimensions, including system architecture (e.g., centralized, distributed, and cloud-native), data models (e.g., relational, graph, time-series, and vector), storage structures (e.g., row store, column store, LSM-tree, and B$^{+}$-tree), and deployment environments (e.g., on-premises, clouds, and containers). These advances enable DBMSs to support a wide variety of workloads such as Online Transaction Processing (OLTP), Online Analytical Processing (OLAP), Hybrid Transactional/Analytical Processing (HTAP), Business Intelligence (BI), and even Machine Learning (ML). Numerous DBMSs have been developed, including commercial systems like Oracle~\cite{oracle}, IBM Db2~\cite{db2}, Microsoft SQL Server~\cite{sqlserver}, Amazon Aurora~\cite{aurora}, Google BigQuery~\cite{bigquery}, and Snowflake~\cite{snowflake}, as well as open-source systems such as PostgreSQL~\cite{postgresql}, MySQL~\cite{mysql}, DuckDB~\cite{duckdb}, ClickHouse~\cite{clickhouse}, Neo4j~\cite{neo4j}, and Apache Spark~\cite{spark}. These systems provide core functionalities for managing structured and semi-structured data, including data storage, querying, indexing, transactional guarantees (ACID), and analytical capabilities, thereby forming a foundational layer for modern industrial and business applications~\cite{stonebraker2024goes}.

Since 2020, Large Language Models (LLMs) have emerged as new transformative technologies with the ability to understand, generate, and reason over unstructured and multimodal data~\cite{zhao2023survey}.
Built on transformer architectures and trained on massive corpora, LLMs exhibit remarkable generalization abilities across diverse tasks with minimal supervision or fine-tuning.
The LLM ecosystem is evolving very rapidly with diverse models. Early language models include GPT-3~\cite{floridi2020gpt}, GPT-4~\cite{achiam2023gpt},  PaLM~\cite{chowdhery2023palm}, and Claude~\cite{claude}. Open-source LLMs such as LLaMA~\cite{touvron2023llama}, Mistral~\cite{mistral}, DeepSeek~\cite{liu2024deepseek, guo2025deepseek}, and Qwen~\cite{bai2023qwen} have further enriched the LLM ecosystem.
Beyond text-based LLMs, Vision-Language Models (VLMs) and Multi-modal LLMs (MLLMs) have also become increasingly prominent, such as Gemini~\cite{team2023gemini, team2024gemini}, GPT-4o~\cite{hurst2024gpt}, CLIP~\cite{clipllm}, Flamingo~\cite{alayrac2022flamingo}, ChatGLM~\cite{glm2024chatglm}, and Kosmos-2~\cite{peng2023kosmos}. Those models integrate visual, textual, and sometimes auditory modalities, enabling advanced reasoning over complex and multimodal inputs.
LLMs can be deployed across various platforms, including cloud-based services, on-device components, or integrated agents, enabling a wide range of intelligent functionalities across industries. As their capabilities continue to expand, LLMs are increasingly regarded as a new layer of general-purpose infrastructure, particularly for intelligent applications that require natural language interaction, semantic reasoning, and multi-modal understanding.

\begin{table*}[!t]
    \centering
    \caption{Comparison of DBMSs and LLMs in industrial and business applications.}
    \scalebox{1}{
\begin{tabular}{ l  l  l }
\hline
\textbf{Aspect} & \textbf{DBMSs} & \textbf{LLMs} \\ \hline

Primary Purpose & Data Storage, Query, Management & Natural Language Understanding, Reasoning, Generation \\ \hline

Main Use & \makecell[l]{OLTP/OLAP/HTAP/BI, Indexing\\ and Querying, Transaction Processing, etc.} & Q\&A , Semantic Search, Text Summarization, etc. \\  \hline

Strengths & \makecell[l]{ACID Guarantees, High Query Efficiency,\\ Data Integrity, Joins and Aggregations, etc.} & \makecell[l]{Human-like Interaction, Unstructured Data Processing,\\ Fine-tuning, Generalization from Examples, etc.} \\ \hline

Data Model/Modality& \makecell[l]{Structured and Semi-Structured (e.g., Relation,\\ Graph, JSON, Time-Series)}  & Unstructured Data (e.g., Text, Documents, Images, Videos) \\ \hline

Interface & SQL, APIs & Natural Language Prompts, APIs  \\ \hline

Interpretability  & Transparent Outputs, Well-understood Plans & Opaque Reasoning, Black-box Behavior \\ \hline

Latency & Typically Low-latency, Optimized Query Plans & Often Higher Latency, Depends on Model and Context Size \\ \hline

Update & Tuple-level Updates, Transactions Support  & No Native Data Update, Retraining or RAG Needed \\ \hline
\end{tabular}
}

    \label{tab:comparison_of_dbms_and_llm}
\end{table*}

Current industrial and business systems are increasingly driven by the \textbf{Data+AI} paradigm, in which DBMSs and LLMs are expected to function as complementary and coexisting \emph{dual infrastructures} to support modern applications. This relationship can be expressed as:
\begin{equation}\label{eq:DualInfras}
\text{LLMs} + \text{DBMSs} \rightarrow \text{Dual Infrastructures of Enterprises}
\end{equation}
In this paradigm, both \emph{data} and \emph{models} are regarded as strategic assets. Table~\ref{tab:comparison_of_dbms_and_llm} summarizes and compares the core roles, capabilities, and characteristics of DBMSs and LLMs. While originating from distinct technical backgrounds and solving different problems, DBMSs and LLMs are not competing technologies. Instead, their integration offers the potential to unlock powerful synergies that surpass the capabilities of either system can achieve in isolation.
There are several motivating factors for integrating DBMSs and LLMs into a unified system. For example, many modern applications require hybrid query processing that combines structured data retrieval (e.g., SQL) with unstructured data understanding, natural language interpretation, or semantic reasoning. 
Another key motivation is system-level synergy, such as leveraging DBMS features (e.g., indexing, caching, and transaction management) to boost the efficiency and consistency of LLM-based operations, particularly in scenarios involving large-scale or dynamic datasets.
Lastly, the growing field of Industrial Large Models (ILMs) increasingly demands systems where DBMSs manage structured backends while LLMs provide interpretability, reasoning, and language interfaces~\cite{zhou2024industrialsurvey, zhou2024industrial}.


The choice of integration strategy significantly affects system performance, complexity, and maintainability. 
It determines how the data flows between DBMSs and LLMs, how the components interact, and how easily the system can adapt to changes.
Integration is not merely about invoking LLM APIs from within a database or vice versa. Rather, it involves deeply embedding and aligning LLMs and DBMSs across different levels of the system stack, such as the system level, component level, and processing pipeline level. Achieving this requires a thorough understanding of the internal architectures, capabilities, and operational semantics of both DBMSs and LLMs.
To meet diverse application needs, various DBMS–LLM integration architectures have been proposed in industry and business, ranging from simple pipeline connectors to deeply integrated strategies. For instance, a simple form of integration involves connecting DBMSs and LLMs via external data pipeline tools, where data flows between components without mutual understanding. A more advanced approach introduces a middleware layer to manage interactions and task decomposition. Deeper strategies embed LLMs directly into the DBMS execution engine as custom query operators or user-defined functions (UDFs). Finally, cloud-native platforms increasingly offer end-to-end integration capabilities, with providers like Oracle, Google, Amazon, and Alibaba launching unified environments where both DBMSs and LLMs can be tightly coupled and jointly optimized.

In this paper, we aim to provide a comprehensive survey of existing DBMS–LLM integration architectures, along with an analysis of key research challenges and open directions in this evolving area. 
Our main contributions are summarized as follows:
\begin{itemize}
\item \textbf{Survey of the Current Status}. We systematically review and categorize existing DBMS-LLM integration approaches in industrial and business applications, covering a wide range of use cases and architectural patterns. We also provide recommendations for choosing the optimal integration strategies according to specific application requirements and system constraints.   
\item \textbf{Analysis of Future Research Challenges}. We also identify and discuss some key challenges in DBMS–LLM integration to enable more robust, scalable, and intelligent integrated systems in the future.
\end{itemize}


\begin{figure*}[!t]
\centering
\includegraphics[width=0.8\linewidth]{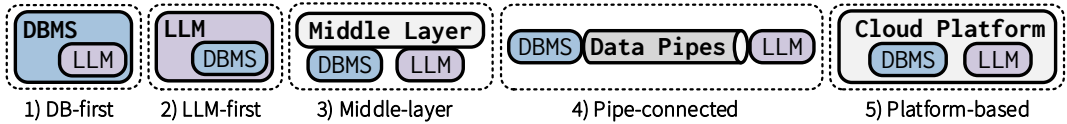}
\caption{Illustrations of different integration strategies.}
\label{fig:illustration_of_integration}
\end{figure*}

\begin{table*}[!ht]
    \centering
    \caption{Comparison of strengths and weaknesses of different DBMS-LLM integration strategies.}
    \scalebox{0.88}{
\begin{tabular}{ l  l  l  l }
\hline
\textbf{Strategy} & \textbf{Features} & \textbf{Strengths} & \textbf{Weaknesses} \\ \hline

DB-First & \makecell[l]{• LLM embedded as a DB component (e.g.,\\ interface, optimizer) or operator (e.g.,\\ UDF into query execution plans) \\ • DB is in full control} & \makecell[l]{• Tight integration with query engines \\• Benefits from database indexing \\• Easier governance} & \makecell[l]{• Limited LLM context window \\• Hard to manage model versions \\• DB constraints limit LLM potential} \\ \hline

LLM-First & \makecell[l]{• DB acts as a machine (e.g., caching \\or indexing component) inside LLM \\• LLM is central to drives logic} & \makecell[l]{• Flexible and user-friendly \\• Natural interface for non-experts \\• Easily handles unstructured data} & \makecell[l]{• Hallucinations or incorrect SQL \\• Weak security/access control \\• DB optimization not utilized well} \\ \hline

Middle-Layer & \makecell[l]{• Orchestration layer mediates between \\DB and LLM by tools like LangChain \\• Intermediary system coordinates \\communication, data flow, pipelines} & \makecell[l]{• Modular design \\• Composable pipelines \\• Easier to evolve components \\• Rich interaction patterns} & \makecell[l]{• Increased system complexity \\• Performance bottlenecks \\• Harder to debug and test} \\ \hline

Pipe-Connected & \makecell[l]{• DB and LLMs run independently and are \\integrated via stream or batch data pipelines \\• Dataflow-centric architecture} & \makecell[l]{• Highly decoupled and modularized \\• Well-suited for event-driven architectures \\• Separation of concerns} & \makecell[l]{• High latency or eventual consistency \\• Complex to manage state and recovery \\• Debugging across stages can be hard} \\ \hline

Platform-Based & \makecell[l]{• Cloud platforms offer LLM+DB as managed \\services (e.g., Snowflake Cortex, Google Cloud,\\  and Oracle Cloud)} & \makecell[l]{• Low setup cost (ready to use) \\• Full-stack scalability \\• Vendor ecosystem support} & \makecell[l]{• Vendor lock-in \\• Limited transparency \\• Hard to customize deeply} \\ \hline
\end{tabular}
}

    \label{tab:comparison_of_integration_strategies}
\end{table*}


\section{Related Work}
A growing body of surveys has explored the intersection between LLMs and DBMSs, reflecting the increasing interest in combining traditional data management with novel reasoning capabilities.

\textbf{Interactions Between DBMSs and LLMs}.  
Several recent studies investigate how LLMs can interact with traditional database systems. Pan et al.~\cite{pan2024unifying} present a roadmap for unifying Knowledge Graphs (KGs) and LLMs, outlining some design principles and use cases that benefit from their integration.
Similarly, Khorashadizadeh et al.~\cite{khorashadizadeh2024research} provide a comprehensive survey of collaborative strategies that leverage LLMs and KGs to enable more advanced and interpretable reasoning.
Kim et al.~\cite{kim2024trustworthy} categorize various DBMS-LLM interaction paradigms, where LLMs serve as data sources, data processors, or query translators between structured and unstructured data.

\textbf{LLM-enhanced Data Management}. LLMs have also been applied to enhance core data management tasks. Hong et al.~\cite{hong2024next} explore the use of LLMs for SQL generation and complex query interpretation, demonstrating the potential of language models in bridging natural language interfaces with structured queries. Zhou et al.~\cite{zhou2024llm} examine the broader role of LLMs in data management tasks such as data cleaning and entity resolution. However, their discussion is largely high-level and lacks a technical taxonomy or system-level analysis of integration strategies.

\textbf{Vector Databases}. Vector databases have become a foundational infrastructure for enabling LLM-based retrieval tasks, particularly in retrieval-augmented generation (RAG) pipelines.
Surveys by Pan et al.~\cite{pan2024survey}, Han et al.~\cite{han2023comprehensive}, and Jing et al.~\cite{jing2024large} provide comprehensive overviews of vector database architectures, indexing methods, and similarity search techniques. The integration of LLMs into these systems to enhance semantic retrieval and reasoning capabilities is also addressed in~\cite{jing2024large}.

\textbf{Databases Meet AI}. In the broader context of database and AI, Zhou et al.~\cite{zhou2020database} provide early insights into the bidirectional relationship of AI4DB and DB4AI. Cai et al.~\cite{cai2022survey} survey the use of deep reinforcement learning (DRL) in data analytics and database operations. Yan et al.~\cite{yan2023join} focus specifically on DRL-based techniques for join order selection, a key challenge in query optimization.

Compared to these works, our paper presents a focused and in-depth survey of architectural-level integrations between LLMs and DBMSs. We categorize and analyze five representative integration patterns, highlight their trade-offs, and identify open challenges to guide future research in this evolving domain.












\section{Current Status of LLM-DBMS Integration Strategies}\label{sec:survey}

In this section, we present a detailed introduction to current strategies for integrating LLMs with DBMSs. We first provide a high-level overview and comparison of integration strategies by summarizing their key features, advantages, and trade-offs. Then, we offer an in-depth discussion of each strategy.

\subsection{Overview of Integration Strategies}
During the early Deep Learning (DL) era (2010s to early 2020s), there emerged a growing demand for integrating DL pipelines with DBMSs. Lixi et al.~\cite{zhou2024serving} summarized three common architectural paradigms: \emph{DL-centric}, \emph{UDF-centric}, and \emph{Relation-centric}. They also proposed a vision for an advanced system architecture that seamlessly integrates these paradigms, along with hybrid designs that bridge the gap between the three paradigms. Their works offer valuable insights for the LLM-DBMS integration in the era of LLMs.

Typically, the integration of LLMs and DBMSs can be characterized along several dimensions: 1) the purposes of integrations (e.g., query understanding, augmentation), 2) the system layer or endpoint where the integration occurs (e.g., inside the DBMS, at the interface, or externally), and 3) the degree and depth of coupling between the LLM and DBMS (e.g., tightly integrated or loosely coupled). 

Based on these dimensions, we identify five major integration strategies, as illustrated in Figure~\ref{fig:illustration_of_integration}. Each strategy reflects a different structural and functional relationship between the DBMS and the LLM. For example, the DB-first strategy embeds the LLM directly within the DBMS, treating it as an internal module or an operator. In contrast, the LLM-first strategy treats the DBMS as an external machine, possibly for executing structured sub-queries generated by the LLM or for accessing specific records during LLM-driven reasoning.
These strategies vary in their architectural focus and operational trade-offs. Table~\ref{tab:comparison_of_integration_strategies} summarizes the comparative strengths and limitations of each strategy.

It is important to note that our categorization is \textbf{non-orthogonal}: some of the integration dimensions are not mutually exclusive, and hybrid combinations frequently occur in real-world industrial and business systems.
For example, a system may simultaneously adopt a DB-first architecture to embed LLM-enhanced execution operators directly within the query engine, while also leveraging pipe-based tools (e.g., Apache Flink) to facilitate data exchange with external LLM services for tasks such as semantic enrichment or user query interpretation.

\subsection{DB-first Strategy: Integration on the DB-side}

\begin{figure}[!t]
\centering
\includegraphics[width=0.65\linewidth]{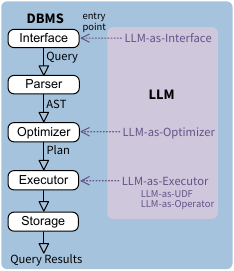}
\caption{The architecture and processing pipelines of DBMSs, as well as the entry points of LLMs.}
\label{fig:DBMS_architecture_and_pipelines}
\end{figure}

The DB-first strategy positions a traditional DBMS as the core architecture of the system, with LLMs serving as auxiliary modules or external services. Figure~\ref{fig:DBMS_architecture_and_pipelines} shows the typical DBMS architecture and processing pipelines that include the following stages: 1) User Interface: accepting user queries like SQL, 2) Query Parsing: converting the query into an Abstract Syntax Tree (AST) or other internal representations, 3) Query Optimization: performing logical and physical query optimization to generate execution plans, and 4) Query Execution: executing physical query plans that are typically represented as a tree or a Directed Acyclic Graph (DAG) of operators.

DB-first strategies can preserve the traditional DBMS processing guarantees while augmenting them with LLM-driven intelligence. In this architecture, the DBMS remains the core system, and LLMs are selectively integrated into specific components or stages of the processing pipeline to enhance functionality. As illustrated in Figure~\ref{fig:DBMS_architecture_and_pipelines}, different entry points exist for such integration, enabling different use cases like \emph{LLM-as-Interface}, \emph{LLM-as-Optimizer}, and \emph{LLM-as-Executor}.



\textbf{LLM-as-Interface}.
Structured query languages such as SQL can be overly complex and unintuitive, even for experienced database experts. One of the most natural and widely adopted ways is to leverage LLMs' natural language understanding capabilities to translate user intents expressed in natural language into executable SQL queries. This line of research and development is commonly referred to as Text-to-SQL (Text2SQL or NL2SQL). Several recent works have contributed significant advances in this area. 
Jinyang et al.~\cite{li2023can} introduce \method{BIRD}, a large-scale Text2SQL benchmark targeting complex real-world databases. \method{BIRD} includes task challenges such as noisy data, external knowledge dependencies, and efficiency constraints. Their findings indicate that even state-of-the-art models like GPT-4 still struggle with these real-world complexities, highlighting the gaps between academic performance and practical usability.
Kaushikpresent et al.~\cite{rajan2024welding} propose \method{NL2Weld}, a system that can translate natural language directly into the Weld Intermediate Representation (IR) using GPT-4. This approach bypasses SQL entirely, enabling more optimized data analytics by mapping natural language to a lower-level, execution-friendly representation.
Tianshu et al.~\cite{Wang2025DBCopilot} present \method{DBCopilot}, a schema routing framework designed to handle large and heterogeneous databases. It efficiently selects relevant schemas for a given natural language query, which can improve the scalability and accuracy of schema-agnostic Text2SQL generation. 
Simone et al.~\cite{papicchio2025think2sql} investigate the impact of various training strategies—such as zero-shot prompting, supervised fine-tuning with reasoning traces, reinforcement learning, and their combinations—on the performance of LLMs in Text2SQL tasks. Their results show that combining fine-tuning and reinforcement learning achieves the best generalization and execution accuracy, allowing smaller models to rival the performance of much larger ones.
Despite these advancements, it is important to note that Text2SQL primarily addresses the user interface layer, transforming how users interact with databases but not altering the core query processing, optimization, or execution logic of the DBMS itself. This is why some researchers argue that focusing solely on Text-to-SQL is insufficient for truly intelligent and adaptive database systems, and deeper integration strategies are needed to unlock the full potential of LLM–DBMS synergy~\cite{biswal2025text2sql}.

\textbf{LLM-as-Optimizer}. The query optimizer is often referred to as the \emph{brain} of a DBMS, as it plays a critical role in determining the most efficient execution plans for a given query. Given their strong reasoning, abstraction, and pattern recognition capabilities, LLMs are increasingly being explored as promising tools to enhance or even replace traditional optimization components, including tasks such as query rewriting, join order enumeration, and execution plan selection.
Zhaodonghui et al.~\cite{li2024llmr2} propose \method{LLM-R2}, a hybrid system that integrates LLMs with traditional rule-based query rewriting to improve SQL execution efficiency while preserving query equivalence. The system employs a contrastive curriculum-trained representation model to select high-quality example rewrites, enabling the LLM to generate more effective and semantically valid query rewrites without relying on potentially inaccurate cost models.
Building upon this idea, Zhaoyan et al.~\cite{sun2024rbot} design \method{R-Bot}, which incorporates multi-source rewrite evidence and a hybrid structure-semantic retrieval mechanism. This architecture grounds the LLM's output in verified rule sets and high-quality Q\&A pairs, thereby reducing hallucinations and improving the robustness and reliability of the rewriting process.
Jie et al.~\cite{tan2025can} present \method{LLM-QO}, a novel framework that directly generates SQL execution plans using LLM's fine-tuning abilities. This approach bypasses traditional plan enumeration and cost modeling altogether. Experimental evaluations show that \method{LLM-QO} consistently outperforms both classical and learned optimizers across diverse datasets, demonstrating strong generalization to unseen queries.
Zhiming et al.~\cite{yao2025query} introduce \method{LLMOpt}, a comprehensive query optimization framework based on fine-tuned LLMs. \method{LLMOpt} combines plan generation with global candidate selection, eliminating reliance on heuristic search or inaccurate cost estimators. Benchmarking results reveal that \method{LLMOpt} delivers superior performance compared to systems such as PostgreSQL and some machine learning based query optimizers.
Nikita et al.~\cite{vasilenko2025training} propose \method{LLM-PM}, a training-free, lightweight framework that leverages LLM-derived embeddings of execution plans. Rather than modifying the DBMS itself, \method{LLM-PM} augments existing cost-based optimizers by suggesting performance hints, such as join strategies or join tree shapes (e.g., left-deep or bushy trees). Through a nearest-neighbor search over plan embeddings and a two-neighborhood consistency check, \method{LLM-PM} achieves significant query runtime improvements on standard benchmarks, validating the utility of plan representation learning for real-world optimization.
In summary, these works demonstrate that LLMs can play an increasingly active role in query optimization—either by generating execution plans directly, augmenting existing optimizers with semantic insight, or automating the traditionally manual process of rule-based rewriting. Unlike LLM-as-Interface approaches, these strategies target the core logic of query processing, pushing the boundaries of intelligent and adaptive DBMS design.

\textbf{LLM-as-Executor}. 
The execution engine of a DBMS is responsible for evaluating physical query plans and generating query results. Within this component, LLMs can be integrated in two primary modes: as User-Defined Functions (UDFs) and as native operators embedded in the query execution pipelines. Both modes enable the system to support advanced semantic and reasoning operations during query execution, such as classification, summarization, and multi-modal inference.
\textbf{(1) LLM-as-UDF}.
UDFs have long served as a powerful mechanism for injecting user-defined logic into SQL-based systems~\cite{sichert2022user, franz2024dear, foufoulas2023efficient}.
The LLM-as-UDF paradigm encapsulates LLM capabilities inside SQL-callable functions, allowing LLM inference to be invoked directly in SQL queries. This approach enables tight coupling between traditional database processing and the reasoning abilities of LLMs, without requiring complex external orchestration.
For example, Parker et al.~\cite{glenn2024blendsql} propose \method{BlendSQL}, which integrates LLMs with SQLite by exposing specialized LLM-powered SQL functions. This allows users to perform semantic mapping and contextual reasoning over tabular data inside the database engine.
Similarly, Fuheng et al.~\cite{zhao2025hybrid} introduce Hybrid Query User-Defined Functions (HQUDFs) that combine structured queries over relational data with semantic inference over unstructured knowledge via LLMs, enabling rich hybrid querying capabilities.
Anas et al.~\cite{dorbani2025beyond} present \method{FlockMTL}, which integrates LLMs directly into the {DuckDB} query engine. \method{FlockMTL} exposes LLM functionality as scalar and aggregate SQL functions, allowing tuple-level or group-level semantic operations. These UDFs can perform language understanding tasks like classification or summarization as part of the query execution process.
\textbf{(2) LLM-as-Operator}.
In contrast to the UDF mechanism, the LLM-as-Operator strategy incorporates LLMs as native operators within the physical query execution pipeline. This approach treats LLMs as first-class citizens of the DBMS, enabling them to participate directly in execution plans and interact with database operators~\cite{passing2017sql}. 
Mohammed et al.~\cite{saeed2023db, Saeed2024QueryingLL} introduce \method{GALOIS}, a system that executes SQL queries on top of pre-trained LLMs. \method{GALOIS} decomposes SQL queries into logical plans, where physical operators are implemented via structured LLM prompts. These prompts guide the LLM to retrieve or synthesize structured outputs from its latent knowledge or textual embeddings, supporting query execution directly over encoded unstructured data. 
Matthias et al.~\cite{urban2024eleet} present \method{ELEET}, which introduces multi-modal operators (MMOps) into the DBMS. These native operators—such as multi-modal scan, join, and union—use small language models (SLMs) as extractive decoders. \method{ELEET} allows structured tuples to be extracted from semi-structured or unstructured sources (e.g., text), enabling multi-row transformations during query execution.
Expanding this concept, Matthias et al.~\cite{urban2024caesura, urban2024demonstrating} develop \method{CAESURA}, a system that uses LLMs as multi-modal query planners. \method{CAESURA} defines four distinct multi-modal operators—\emph{VisualQA}, \emph{TextQA}, \emph{Python UDF}, and \emph{Image Select}—to support unified querying across diverse data modalities including text, images, and code. The LLM provides coordination and semantic interpretation within the physical query plan, enabling flexible and expressive multi-modal queries.

\subsection{LLM-first Strategy: Integration on the LLM-side}
In contrast to the DB-first strategy, which embeds LLM functionalities within the database system, the LLM-first strategy positions the LLM as the central orchestration layer. In this architecture, the database is treated as a supporting backend component, often used for storage, caching, or structured retrieval, while the LLM handles the primary logic, reasoning, and interaction tasks. It is especially suitable for applications where LLMs serve as the primary user interface or analytical engine, such as conversational agents or intelligent assistants.
Compared with DB-first strategies, integration options of LLM-first strategies are more limited on the LLM side, since traditional DBMSs are not designed to be directly embedded within LLMs. Instead, DBMSs are often used as retrieval backends or cache layers to accelerate query responses or support Retrieval-Augmented Generation (RAG) pipelines.

Wenbo et al.~\cite{Sun2025DatabaseIA} introduce \method{TranSQL}, a toolkit that integrates relational databases directly within LLMs by translating neural network operations into SQL queries and storing model weights as relational tables. This allows a relational database to efficiently manage LLM execution using built-in database features like disk-to-memory management and caching. \method{TranSQL} enables end-to-end transformer-based text generation entirely within a relational environment, eliminating the need for specialized deep learning infrastructure.
Alekh et al.~\cite{jindal2024turning} present \method{GOD Machine}, an ambitious framework that turns relational databases into generative AI systems. It integrates LLMs with logical data models (LDMs) of databases to provide scalable, interpretable, and privacy-preserving analytics. This system automates data modeling, retrieval, and pipeline management through scalable retrieval mechanisms and introduces a framework for building end-to-end AI-powered applications directly on relational databases.
Xinyang et al.~\cite{zhao2024chat2data} develop \method{Chat2Data}, which is an interactive data analysis system that combines LLMs, vector databases, and RAG techniques. This system supports natural language querying across both structured and unstructured data. A key component of the architecture is a vector database used as a cache layer, storing embeddings of frequently asked questions and their corresponding answers to improve system responsiveness and reduce redundancy.

The LLM-first strategy emphasizes semantic abstraction, natural interaction, and integration simplicity, enabling rapid development of AI-powered applications without tightly coupling LLMs to low-level database internals. However, this approach often sacrifices fine-grained query control and data processing guarantees, making it less suitable for mission-critical transactional workloads or performance-sensitive analytical queries. In practice, LLM-first strategies are commonly used for user-facing interfaces, exploratory data analysis, and knowledge-driven reasoning tasks.

\subsection{Middle-layer Strategy: Integration via a Middleware}


Distinct from the DB-first and LLM-first paradigms that treat either the database or the LLM as the central component, the Middle-layer strategy introduces an intermediary orchestration layer that coordinates the two systems and processing pipelines.
This middleware sits between the LLM and the DBMS, acting as a smart controller to manage, optimize, and unify the processing pipelines of both components. 

The key motivation behind this strategy is to bridge two fundamentally different data processing paradigms: 1) the deterministic, rule-based query and retrieval pipeline of a traditional DBMS, and 2) the stochastic, probabilistic generation and reasoning pipeline of an LLM. Therefore, the essence of integration lies in building a middleware that can compose, schedule, and optimize hybrid tasks, allowing smooth and efficient interaction between structured queries and semantic reasoning.

Yu et al.~\cite{gu2024middleware} propose equipping LLMs with customized middleware tools that serve as an intermediate layer between LLMs and external systems such as knowledge bases or relational databases. These tools interpret user intent, generate executable queries (e.g., SQL), and return results to the LLM, which mitigates token-length limitations and improves task performance in complex environments.
Sumedh et al.~\cite{rasal2024multi} presents a multi-LLM orchestration engine that integrates multiple LLMs with a temporal graph database and a vector database to enable personalized, context-rich AI assistance. The temporal graph database captures evolving conversational history and user preferences over time, while the vector database securely encodes private data for precise retrieval.
The middleware coordinates these components to deliver personalized, privacy-preserving, and context-rich responses, without retraining the LLMs. This addresses major challenges such as long-term memory, hallucination, and private data integration.
Jiayi et al.~\cite{wang2025aop} propose \method{AOP}, a middleware framework 
that unifies LLM-based semantic reasoning with DBMS query processing. It dynamically composes execution pipelines containing both semantic operators (invoked via LLMs) and structured query operators (e.g., SQL or dataframe operations). By integrating over heterogeneous data sources in data lakes, \method{AOP} orchestrates hybrid workflows across unstructured and structured data, achieving flexibility and performance gains.

\subsection{Pipe-connected Strategy: Integration via Data Pipes}
Pipe-connected strategies refer to integrating LLMs and DBMSs using data pipe connectors or tools, which is a pragmatic, loosely coupled approach that leverages mature streaming and ETL technologies to enable real-time or batch data exchange. Unlike DB-first or LLM-first strategies that embed functionality into the system core, the pipe-connected approach preserves system modularity by treating the LLM and DBMS as independent services connected via well-defined data flows.
This integration is particularly effective in event-driven architectures or scenarios requiring scalable, real-time analytics, where tight coupling may not be necessary or practical. Modern data processing tools (e.g., Apache Kafka~\cite{kafka}, Apache Flink~\cite{flink}, and Spark Streaming~\cite{spark}) enable high-throughput, low-latency pipelines, facilitating seamless integration between components without requiring deep architectural fusion.

Moreh et al.~\cite{park2024dataverse} propose \method{Dataverse}, which is an open-source, user-friendly ETL platform designed specifically for preparing large-scale datasets for LLM training and inference. It features a block-based workflow interface and supports scalable data processing via integration with systems like Apache Spark and AWS EMR. Key features include data deduplication, decontamination, and bias mitigation. \method{Dataverse} can serve as a bridge between data stored in DBMSs and LLMs by enabling automated ingestion, transformation, and delivery of curated datasets for semantic tasks.
Kai et al.~\cite{waehner2025ultimate} position data streaming as a transformative paradigm that enables real-time decision-making and event-driven innovation across industries. They emphasize organizational strategies (e.g., fostering internal streaming communities) and showcase emerging trends such as integrating LLMs into streaming pipelines for enhanced reasoning and semantic analytics. \cite{waehner2025ultimate} also gives a case about an autonomous airport, which leverages data streaming, DBMSs, and LLMs in its digital transformation to enhance airport operations and passenger experiences.

Overall, the pipe-connected strategy offers several practical benefits like scalability, flexibility, and ease of integration. However, its loose coupling also introduces several challenges, such as latency overhead in complex pipelines, and increased engineering effort for managing failure recovery.

\subsection{Platform-based Strategy: Integration within a Cloud Platform}

The platform-based strategy focuses on integration modes enabled by cloud-native platforms, where both LLMs and DBMSs are deployed and managed as modular services in the cloud. This category emphasizes deployment architecture and interoperability capabilities provided by modern platforms, rather than the functional embedding of one system into another.

Over the past decades, cloud-native technologies have matured to support highly flexible service delivery modes, such as Software-as-a-Service (SaaS) for DBMSs, Model-as-a-Service (MaaS) for LLMs, and even Platform-as-a-Service (PaaS) for the whole application. These services are typically built on microservices and serverless architectures, which now form the backbone of IT infrastructure in modern enterprises. According to Gartner, over half of enterprise databases are now cloud-hosted, with a growing percentage being fully cloud-native~\cite{dong2024cloud}. Leading cloud providers, including Amazon AWS, Google Cloud, Microsoft Azure, Oracle Cloud, IBM Cloud, Alibaba Cloud, Tencent Cloud, and Huawei Cloud, offer rich ecosystems that include both LLM and DBMS services. This create significant opportunities for integrating LLMs and DBMSs within the same cloud platform, enabling unified resource management, coordinated data processing pipelines, standardized communications, and orchestrated multi-component workflows.

A unified cloud platform can act as a communication and control hub where both LLMs and DBMSs are able to exchange data and participate in joint pipelines. For instance, cloud-native services like AWS Lambda, Azure Logic Apps, or Google Cloud Workflows can coordinate the invocation of LLM models and SQL-based data processing in a seamless, event-driven fashion. Recent developments have also popularized workflow-based architectures that use LLMs as autonomous agents to drive decision-making and database interactions. These workflows support multi-step task execution, retrieval-augmented generation, data validation, and response generation in a unified loop. This is consistent with trends in cloud-native data mesh architectures~\cite{Plazotta2024Data} and agent-based LLM orchestration frameworks. For example, Snowflake Cortex integrates LLM capabilities directly into its data cloud, allowing users to perform LLM-powered analysis on structured data using SQL-like interfaces. Oracle Cloud offers integrated solutions where Oracle Autonomous Database and Oracle LLM Services work together under a shared compute and storage layer to support intelligent workflows.

In summary, platform-based strategies abstract away infrastructure concerns and allow developers to compose and orchestrate complex DBMS–LLM pipelines using standardized cloud-native tools and workflows.

\subsection{What Strategies to Choose?} 
\begin{table*}[!ht]
    \centering
    \caption{Feature-oriented comparison of DBMS-LLM integration strategies (\strong = Strong, \moderate = Moderate, \weak = Weak).}
    \scalebox{1}{
\begin{tabular}{ l  c  c  c  c c}
\hline
\textbf{Dimension} & \textbf{DB-First} & \textbf{LLM-First} & \textbf{Middle-Layer} & \textbf{Pipe-Connected} & \textbf{Platform-Based}\\ \hline

Coupling Degree & \strong & \weak & \moderate & \weak & \moderate \\ \hline

Real-Time & \strong & \weak & \moderate & \moderate & \strong \\ \hline

Scalability & Limited by DB & Limited by LLM & \strong  & \strong & \strong  \\ \hline

Extensibility & \weak & \strong & \strong & \strong & \strong  \\ \hline

Complexity & \weak & \weak & \strong & \moderate & \weak  \\ \hline

Security & \strong & \weak & \moderate & \moderate & \strong  \\ \hline

\end{tabular}
}

    \label{tab:comparison_of_integration_strategies_feature_matrix}
\end{table*}
Table~\ref{tab:comparison_of_integration_strategies_feature_matrix} provides a feature-based comparison of the five DBMS–LLM integration strategies, using qualitative scores (Strong/Moderate/Weak) across several dimensions. Based on this analysis, we argue that the optimal integration strategy should be selected according to the following key factors:

\textbf{1) Task Requirements}: The nature of the workloads (e.g., batch vs. streaming, offline processing vs. real-time interaction) strongly affects integration needs. Real-time and low-latency tasks are better served by DB-first, pip-connected, or platform-based integrated solutions, while offline analytics can afford more flexibility in architectural choice.

\textbf{2) Functional Requirements}: If an application demands complex multi-modal or cross-model interactions (e.g., querying both relational tables and unstructured text, or linking visual data with structured knowledge graphs), DB-first integration is advantageous, as it enables native access and processing capabilities over heterogeneous data sources.

\textbf{3) LLM Dependencies}: In scenarios where LLMs heavily rely on advanced LLM functionalities (e.g., code generation, reasoning, instruction following), the LLM-first integration provides greater flexibility in function execution and adaptability. By contrast, the DB-first integration may be constrained by the predefined execution models of database engines.

\textbf{4) Deployment Modes}: When using cloud-native platforms or workflow-based orchestration systems, platform-based integration offers significant benefits, including unified management of both data and model services, simplified scaling, and standardized APIs for cross-component communications.

\textbf{5) Hybrid Strategies}: In some practical settings, hybrid integration strategies may offer the best trade-offs. For example, combining DB-first integration for data access efficiency with platform-based coordination for scalability and flexibility allows systems to balance performance with manageability.







\section{Future Challenges}\label{sec:challe}

In this section, we summarize some key open challenges in the integration of DBMSs and LLMs, with a focus on architectural, algorithmic, and performance aspects.

\subsection{Challenges in DB-First Strategies}

In DB-first strategies, the DBMS acts as the primary control system, embedding LLMs as logical operators or external functions. Several challenges arise from this tightly coupled integration, such as:

\begin{itemize}
    \item \textbf{Cross-Model Query Execution.} Systems like \cite{urban2024efficient} propose extending query processing to include both relational and unstructured text data. A major challenge is designing unified execution models and query languages that can handle hybrid data types (e.g., structured tables, semi-structured graphs, unstructured text, and images) efficiently. For instance, \method{HybridGraph}~\cite{ammar2025towards} introduces graph-based models that mix relational and semantic representations, but efficient execution, indexing, and optimization remain underexplored.

    \item \textbf{Cost Modeling and Plan Optimization.} Injecting LLMs into DBMSs' original query processing pipelines will introduce new types of operators with non-deterministic behavior, high latency, and variable computational costs. Traditional cost-based optimizers of DBMSs are ill-equipped to handle these uncertainties. Therefore, how to estimate the cost of the injected LLM-based operators and integrate them into join ordering and plan enumeration frameworks is an open research direction~\cite{wu2020note}.

    \item \textbf{Cross-Modal Operator Design.} As LLMs become increasingly multi-modal, there is a need to define new operator classes for sub-tasks across modalities (e.g., \textit{Image Filter}, \textit{Video Summary}, \textit{Text Extraction}~\cite{urban2024caesura}) and design cross-modal join operators (e.g., joining table rows with images or videos). These operators can be seen as LLM-backed UDFs and must be systematically integrated into the DBMS pipeline. Similar to prior efforts on hybrid relational-graph operators such as in \method{GRFusion}~\cite{hassan2018grfusion}, the challenge is to balance expressiveness, performance, and composability in these new operator designs.
\end{itemize}

\subsection{Challenges in LLM-First Strategies}

In LLM-first architectures, the LLM is the central component, and the DBMS is typically abstracted as a tool (e.g., KV store, SQL plugin, or retrieval engine). This design simplifies end-user interfaces but introduces unique concerns such as:

\begin{itemize}
    \item \textbf{Limited Declarativity and Control.} Delegating control to the LLM sacrifices the optimization capabilities of DBMSs. It becomes difficult to ensure query correctness, efficiency, or even completeness, especially when dealing with complex queries or large data volumes.

    \item \textbf{Tool Use and Reliability.} Prompt-based access to DBMSs introduces tool-call fragility, dependency on context window limits, and inconsistency in tool invocation. Improving tool-calling reliability remains challenging.
\end{itemize}

\subsection{Challenges in Middle-Layer Orchestration}

Middleware-based orchestration strategies act as intermediaries between DBMSs and LLMs. These orchestration layers mediate data access, plan decomposition, operator execution, and caching. However, several challenges still remain, such as:

\begin{itemize}
    \item \textbf{System Coordination.} Coordinating execution between DBMS engines and LLMs requires advanced runtime environments that can manage state, partial results, operator delegation, and errors.
    
    \item \textbf{Dynamic Adaptation.} Middleware systems need to adapt execution strategies based on workload characteristics, latency budgets, and application intent (e.g., information retrieval vs. transaction processing).
\end{itemize}

\subsection{System Design and Application Suitability}

Given the wide range of available integration strategies, a key research question is how to select the most suitable architecture for a given application and workload. As discussed in \cite{Papotti2024Querying}, this is a fundamental decision that depends on latency requirements, data volume, security needs, and user expertise.

\begin{itemize}
    \item \textbf{Quantitative Evaluation Frameworks.} A standardized benchmarking framework is needed to evaluate the trade-offs among strategies in terms of latency, accuracy, cost, interpretability, and robustness.

    \item \textbf{Integration Selection Agent.} Future systems could benefit from meta-controllers or learning-based agents that can dynamically select and configure the integration architecture based on application goals and context.

    \item \textbf{Hybrid Architectures.} Designing and managing hybrid architectures that combine multiple integration strategies poses unique challenges. This includes making principled trade-offs across dimensions such as performance, modularity, fault isolation, and development complexity.
\end{itemize}



    


\section{Conclusion}\label{sec:conclu}



The integration of LLMs and DBMSs represents a promising paradigm shift in the design of intelligent \emph{Data+Model} driven applications. In this paper, we examine current integration strategies and categorize five representative architectural patterns, each offering unique trade-offs in terms of flexibility, performance, and deployment complexity. While these integration efforts unlock new possibilities by bridging structured data processing with unstructured language understanding, they also introduce several open challenges, such as performance optimization and system interoperability.
 Looking forward, future research should focus on developing adaptive and composable integration frameworks, establishing standardized benchmarks for evaluation, and designing robust architectures that can operate reliably at scale. We believe that addressing these challenges is essential for realizing the full potential of synergy between DBMSs and LLMs, which paves the way to the next generation of intelligent, scalable, and trustworthy systems.





\end{document}